\documentclass[a4paper]{article}
\usepackage[T1]{fontenc}
\usepackage[utf8]{inputenc}
\usepackage{amsmath}
\usepackage{amsthm}

\makeatletter

\newcommand*\LyXZeroWidthSpace{\hspace{0pt}}

\numberwithin{equation}{section}
\numberwithin{figure}{section}
\theoremstyle{plain}
\newtheorem{thm}{\protect\theoremname}
\theoremstyle{plain}
\newtheorem{ax}[thm]{\protect\axiomname}
\theoremstyle{plain}
\newtheorem{lem}[thm]{\protect\lemmaname}
\newtheorem{remark}{Remark}

\@ifundefined{date}{}{\date{}}

\usepackage{color}
\usepackage{amsfonts}\usepackage{amsthm}\usepackage{amscd}\numberwithin{equation}{section}
\usepackage{bm}
\usepackage[all]{xy}
\usepackage{hyperref}

\usepackage{algorithm}
\usepackage{algorithmicx}
\usepackage{algpseudocode}
\usepackage[most]{tcolorbox}
\tcbuselibrary{listingsutf8}
\usepackage{float}
\usepackage{lipsum}  
\tcbset{
  highlight algorithm/.style={
    colback=gray!10,
    colframe=black,
    boxrule=0.5pt,
    arc=2pt,
    left=6pt,
    right=6pt,
    top=6pt,
    bottom=6pt,
    width=\textwidth,
    sharp corners,
  }
}

\makeatother

\usepackage[english]{babel}
\providecommand{\axiomname}{Axiom}
\providecommand{\lemmaname}{Lemma}
\providecommand{\theoremname}{Theorem}

\begin{document}
\begin{flushleft}
\textbf{OUJ-FTC-}18\\
\textbf{OCHA-PP-}382\\
 
\par\end{flushleft}

\begin{center}
{\LARGE{}Nambu Non-equilibrium Thermodynamics:}\\
{\LARGE  Axiomatic Formulation and Foundation}{\LARGE\par}
\par\end{center}

\begin{center}
\vspace{32pt}
\par\end{center}

\begin{center}
So Katagiri$^{\dagger1,\dagger2,*}$\footnote{So.Katagiri@gmail.com},
Yoshiki Matsuoka$^{\dagger1}$, and Akio Sugamoto$^{\dagger3}$
\par\end{center}

\begin{center}
\textit{$^{\dagger1}$Nature and Environment, Faculty of Liberal Arts, The Open University of Japan, Chiba 261-8586, Japan}
\par\end{center}

\begin{center}
\textit{$^{\dagger2}$Region of Electrical and Electronic Systems Engineering, Ibaraki University, Nakanarusawa-cho, Hitachi-shi 316-8511, Japan}
\par\end{center}

\begin{center}
\textit{$^{\dagger3}$Department of Physics, Graduate School of Humanities and Sciences, Ochanomizu University, 2-1-1 Otsuka, Bunkyo-ku, Tokyo 112-8610, Japan}
\par\end{center}
\begin{abstract}
We present a theoretical framework for non-equilibrium thermodynamics, termed Nambu Non-equilibrium Thermodynamics (NNET), which unifies reversible dynamics described by the Nambu bracket and irreversible processes driven by entropy gradients.
The formulation provides a covariant description of systems far from equilibrium, where entropy may transiently decrease as a result of reversible circulations or exchanges with the surroundings, extending the applicability of conventional thermodynamic formalisms.

As an illustrative example, a triangular chemical reaction system is analyzed.
It is shown that, without assuming detailed balance or linearity, two geometric structures that behave as conserved quantities in the reversible limit naturally emerge: one associated with cyclic symmetry in the reaction space, and another that vanishes under symmetric reaction rates.
These results demonstrate that NNET provides a unified and covariant formulation for describing both cyclic dynamics and dissipative processes within a single theoretical structure.
\end{abstract}

\section{Introduction}

Dissipation refers to a process in which certain quantities irreversibly decrease as time progresses.
Because of this irreversible nature, dissipative phenomena fundamentally differ from conventional dynamical systems that possess time-reversal symmetry.
In dynamical systems, the number of conserved quantities reflects the underlying symmetries, and the phase space exhibits a kind of rigidity due to the invariance of area under canonical transformations governed by the Poisson bracket.
In short, dissipation characterizes relaxation toward equilibrium, whereas dynamics governed by conservation laws typically represent rotational or cyclic behavior.
\\

Non-equilibrium thermodynamics has traditionally focused on dissipative processes.
Onsager \cite{Onsager_1931,Onsager_1931_2} pioneered a comprehensive formulation of non-equilibrium thermodynamics based on linear response theory, incorporating entropy and transport coefficients.
However, the validity of this formulation relies critically on two assumptions: proximity to equilibrium and the principle of detailed balance.
As a result, the theory is essentially limited to describing dissipative processes that drive systems toward equilibrium.
\\

A central contribution of Prigogine's work was the introduction of entropy flow as a fundamental concept in non-equilibrium thermodynamics \cite{Glansdorff_1964,Kondepudi_2014}.
Open systems—those involving matter exchange with their surroundings—naturally provide a setting for describing physical processes.
Prigogine rigorously separated entropy production from entropy flux and formulated the concepts of non-equilibrium steady states and the minimum entropy production theorem.
He also proposed the General Evolution Criterion (GEC), which decomposes the time variation of entropy production into contributions from changes in thermodynamic forces and flows, ensuring that the part associated with forces is non-positive.
Although this was a remarkable achievement applicable even in nonlinear regimes, it did not provide detailed information about the concrete time evolution of the system.
\\

\color{black}
The GENERIC framework \cite{_ttinger_1997,_ttinger_2005,Grmela_1997,Grmela_2018} later introduced a unified formalism that integrates reversible and irreversible dynamics. Its innovation lies in embedding both components within a thermodynamically consistent geometric structure based on an energy functional, an entropy functional, a Poisson structure, a dissipative structure, and the associated degeneracy conditions.

However, thermodynamic consistency alone does not automatically provide a constructive mechanism for the emergence of far-from-equilibrium organization. In particular, it does not by itself explain how cyclic motion, nonlinear response, pattern formation, or chaotic dynamics arise from the coexistence and competition of conservative circulation, entropy production, and entropy flux in open systems. Thus, even after GENERIC, the constructive problem left open by Prigogine's program remains: how can one formulate concrete macroscopic evolution equations in which circulation-generating and dissipative structures coexist on the same thermodynamic state space?

The absence of such a constructive theory capable of systematically treating coexisting circulation-generating and dissipative structures—and thereby addressing pattern formation and nonlinear responses far from equilibrium—remains an open challenge.
\color{black}
\\

In this paper, we propose a framework called Nambu Non-equilibrium Thermodynamics (NNET), which covariantly integrates reversible structures described by the Nambu bracket and irreversible structures driven by entropy gradients.
The Nambu bracket, first introduced by Yoichiro Nambu in 1973 \cite{Nambu_1973}, is an n-ary generalization of the Poisson bracket defined as a fully antisymmetric Jacobian:

\begin{equation}
\{A_{1},\dots,A_{N}\}\equiv\epsilon^{i_{1}\dots i_{N}}\frac{\partial A_{1}}{\partial x^{i_{1}}}\dots\frac{\partial A_{i_{N}}}{\partial x^{i_{N}}}
\end{equation}

This structure captures the multivariable Jacobian nature of state functions and represents rotations along multiple conserved quantities.
Because it satisfies volume preservation (a generalized Liouville theorem) and allows for the simultaneous conservation of $N-1$ Hamiltonians, it provides a natural description of cyclic dynamics such as those in rigid body rotation or Lotka--Volterra systems\cite{Frachebourg_1996}.
\\

The present paper develops an axiomatic formulation of Nambu Non-equilibrium Thermodynamics (NNET), in which the Nambu bracket is embedded within a hierarchical structure of conserved quantities, enabling a covariant treatment of both dissipative and reversible flows.
The main objectives are as follows:
\begin{enumerate}
	\item	To compare the proposed theory with Onsager's, Prigogine's, and the GENERIC frameworks.
	\item	To formulate an axiomatic structure of non-equilibrium thermodynamics based on the Nambu bracket.
	\item	To clarify the relationship between the proposed theory and the GENERIC approach.
	\item	To demonstrate the formulation using a triangular reaction system as an illustrative example.
\end{enumerate}

\color{black}
In this paper, the scalar function $S$ is referred to as entropy,
in the generalized sense that it generates the irreversible dynamics;
it coincides with the physical thermodynamic entropy
only under appropriate limiting assumptions
such as near-equilibrium regimes.
\color{black}

\section{Axiomatic Formulation of Non-equilibrium Thermodynamics Based on
the Nambu Bracket}

In this section, we present an axiomatic formulation of non-equilibrium
thermodynamics---referred to as Nambu Non-equilibrium Thermodynamics
(NNET)---based on the Nambu bracket\cite{Nambu_1973} . We begin
by introducing the thermodynamic state space.
\begin{ax}
Thermodynamic State Space

The thermodynamic state space $\mathcal{M}$ is a manifold that can
be locally described using a coordinate system composed of thermodynamic
state variables $x^{i}$ with $i=1,\dots,N$.
\end{ax}

Within this space, the subset $\mathcal{M}_{\mathrm{reg}}\subset\mathcal{M}$
denotes the regular region where structures such as time evolution
and entropy can be consistently defined. The axiomatic construction
described below is assumed to hold within $\mathcal{M}_{\mathrm{reg}}$.

Outside $\mathcal{M}_{\mathrm{reg}}$, the definitions of conserved
quantities and entropy may fail to hold. Such points are referred
to as thermodynamic singularities. At these singularities, the uniqueness
of time evolution may break down, and chaotic or aperiodic behavior
can emerge.

We next introduce the dynamical law governing the time evolution of
thermodynamic state variables.

\begin{ax}
Thermodynamic Time Evolution

The time evolution of a thermodynamic state variable $x^{i}$ is expressed
as the sum of a reversible part $\partial_{t}^{(H)}x^{i}$ and an
irreversible part $\partial_{t}^{(S)}x^{i}$:

\begin{equation}
\dot{x}^{i}=\partial_{t}^{(H)}x^{i}+\partial_{t}^{(S)}x^{i}.
\end{equation}
\end{ax}

\begin{ax}
Reversible Part

The reversible part $\partial_{t}^{(H)}x^{i}$ is described by the
Nambu bracket involving $N-1$ Hamiltonians $H_{1},\dots,H_{N-1}$:

\begin{equation}
\partial_{t}^{(H)}x^{i}=\{x^{i},H_{1},\dots,H_{N-1}\}.
\end{equation}
\end{ax}

\LyXZeroWidthSpace{}
\begin{ax}
Irreversible Part

The irreversible part $\partial_{t}^{(S)}x^{i}$ is proportional to
the gradient of a single scalar function, referred to as the “entropy”
$S$, with the proportionality coefficient given by $L_{ij}$. It is
assumed that $L_{ij}$ is positive definite (ensuring convexity of the entropy production term)
):

\begin{equation}
\partial_{t}^{(S)}x^{i}=L^{ij}\frac{\partial S}{\partial x^{j}}.
\end{equation}
\end{ax}

In this formulation, we introduce a scalar potential $S$ that plays the role of an entropy function generating the irreversible dynamics.
It coincides with the physical entropy only under the assumptions of Onsager's linear theory.

Throughout the remainder of this paper, the raising and lowering of
indices for $L^{ij}$ will be omitted for notational simplicity.
\\

\begin{lem}
Conservation of Hamiltonians

In the absence of the irreversible part, the Hamiltonians $H_{1},\dots,H_{n-1}$
are conserved.
\end{lem}

When the irreversible contribution vanishes, the system evolves according
to Nambu mechanics:
\begin{equation}
\dot{x}^{i}=\{x^{i},H_{1},\dots,H_{N-1}\}.
\end{equation}

From this, for any $m\in\{1,2,\dots,N-1\}$ the time evolution of
$H_{m}$ is given by
\begin{equation}
\dot{H}_{m}=\{H_{m},H_{1},\dots,H_{N-1}\}.
\end{equation}

Due to the complete antisymmetry of the Nambu bracket, it follows
that
\begin{equation}
\dot{H}_{m}=0.
\end{equation}

\begin{lem}
Entropy Production

In the absence of the reversible part, the time evolution of the entropy
$S$ is strictly non-negative.
\end{lem}

Indeed, when the reversible contribution vanishes, the entropy evolves
as
\begin{equation}
\dot{S}=L^{ij}\frac{\partial S}{\partial x^{i}}\frac{\partial S}{\partial x^{j}}.
\end{equation}

Since $L_{ij}$ is assumed to be positive definite (i.e., convex),
the entropy production is always positive.

\begin{lem}
Entropy Decrease

When the reversible part is present, the time evolution of entropy
$S$ may become negative.
\end{lem}

\begin{equation}
\dot{S}=\partial_{t}^{(H)}S+\partial_{t}^{(S)}S=\{S,H_{1},\dots,H_{n-1}\}+L^{ij}\frac{\partial S}{\partial x^{i}}\frac{\partial S}{\partial x^{j}}.
\end{equation}

Since the first term is not sign-definite, it is possible for the
total entropy change $\dot{S}$ to be negative.

As Prigogine also pointed out, this corresponds naturally to the entropy flux out of the system into the environment.
\\

We now turn to the necessary and sufficient conditions for a non-equilibrium
steady state.
\begin{lem}
Non-equilibrium Steady State

A state $x_{*}\in\mathcal{M}_{\mathrm{reg}}$ is a complete non-equilibrium
steady state---i.e., $\dot{x}^{i}=0$ for all $i$ ---if and only
if the following condition holds for every $i$:

\begin{equation}
\{x_{*}^{i},H_{1},\dots,H_{N-1}\}=-L^{ij}\partial_{x_{*}^{i}}S.
\end{equation}
\end{lem}

This condition follows straightforwardly from Axioms 2, 3, and 4.
From Lemma 8, the following specific properties can be deduced.
\begin{lem}
Properties of Non-equilibrium Steady States
\end{lem}

\begin{enumerate}
\item \textsl{In the absence of the irreversible part $(L=0)$, a non-equilibrium
steady state is characterized by the vanishing of the Nambu bracket.}
\item \textsl{In the absence of the reversible part (i.e., when the Nambu
bracket vanishes), the steady state condition requires that the entropy
gradient $\frac{\partial S}{\partial x^{i}}$ vanishes.}
\item \textsl{There exist non-equilibrium steady states in which the reversible
and irreversible components exactly cancel each other.}
\end{enumerate}
We now describe a property that becomes particularly important when
analyzing periodic motion.
\begin{lem}
Dissipative Variation of the Hamiltonian

In general systems that include an irreversible part, the Hamiltonian
evolves under the influence of entropy dissipation according to the
relation:
\begin{equation}
\dot{H}_{m}=L^{ij}\frac{\partial H_{m}}{\partial x^{i}}\frac{\partial S}{\partial x^{j}}.
\end{equation}
As a consequence:
\end{lem}

\begin{enumerate}
\item \textsl{$H_{m}$ is conserved if its gradient is orthogonal to the
gradient of the entropy $S$.}
\item \textsl{if $L_{ij}$ is symmetric and positive definite, the direction
of change of $H_{m}$ depends on the relative orientation of the gradients
of $H_{m}$ and $S$: the rate is negative if the two gradients point
in opposite directions, and positive if they are aligned.}
\end{enumerate}
\color{black}
Therefore, in the presence of entropy-driven dissipation, the Hamiltonians are not necessarily strictly conserved quantities. Rather, they function as geometric structures that organize the reversible circulation within the full dynamics.
\color{black}
\\

Together with Lemma 7, this result provides a distinctive feature
of Nambu non-equilibrium thermodynamics and serves as a guiding criterion
for analyzing systems exhibiting periodic oscillations or spiking
behavior.

\color{black}
\begin{remark}[Application to Isolated Systems]
While NNET is particularly powerful for describing open systems where entropy can transiently decrease, the framework naturally encompasses isolated systems. For an isolated system, the total internal energy $E$ must be strictly conserved, and the entropy $S$ must be monotonically non-decreasing. Within NNET, this is naturally achieved by identifying one Hamiltonian with the energy ($H_1 = E$) and another with the entropy itself ($H_2 = S$). Due to the completely antisymmetric nature of the Nambu bracket, the reversible part strictly conserves both energy and entropy ($\partial_t^{(H)} E = 0$ and $\partial_t^{(H)} S = 0$). Furthermore, as deduced from Lemma 10 , the total energy remains strictly conserved under the dissipative dynamics provided that its gradient is orthogonal to the entropy gradient with respect to the transport matrix (i.e., $L^{ij}\frac{\partial E}{\partial x^i}\frac{\partial S}{\partial x^j} = 0$). Under these specific geometric constraints, the total entropy production is governed solely by the irreversible part and is strictly non-negative ($\dot{S} = L^{ij}\frac{\partial S}{\partial x^i}\frac{\partial S}{\partial x^j} \ge 0$).Importantly, the condition $H_2 = S$ clarifies the relationship between NNET and the GENERIC framework in a broad sense. In GENERIC, the requirement that entropy remains invariant under reversible dynamics is structurally imposed as a degeneracy condition (entropy is a Casimir of the Poisson bracket). In NNET, this condition is not rigidly imposed; rather, the isolated-system (GENERIC-like) behavior emerges as a specific geometric choice ($H_k = S$), whereas open-system dynamics allow for $H_k \neq S$.
\end{remark}
\color{black}

\section{Relationship Between the Proposed Framework and GENERIC}

\color{black}
Before comparing NNET with GENERIC, let us briefly clarify the relation between Nambu and Hamiltonian descriptions. In general, it is not known that every Nambu system can be rewritten as an ordinary Hamiltonian system with a single Hamiltonian. Although certain classes of multi-Hamiltonian systems admit Poisson-type decompositions, this does not imply a reduction to a standard Hamiltonian dynamics in the usual sense\cite{Takhtajan_1994}. From the viewpoint of NNET, it is in fact more natural to keep multiple conserved quantities explicit, since the nondissipative part is constructed locally from the Helmholtz decomposition and Darboux theorem directly on the given macroscopic state space\cite{NNET-2}.
\color{black}
\\

\color{black}
In this sense, NNET should not be regarded merely as “GENERIC without degeneration constraints.” If GENERIC is interpreted broadly as a framework combining reversible and irreversible dynamics, then there is certainly a formal similarity. \color{black}However, in a narrower structural sense, the two frameworks differ in how the reversible geometric structure is introduced. In extended-variable or contact-geometric formulations of GENERIC, one may work on an enlarged space involving additional conjugate or flux variables, whereas the Nambu structure in NNET is introduced directly on the original macroscopic state space. Thus, the difference is not only whether degeneration constraints are imposed,
but also how the reversible geometric structure itself is introduced in practice.\color{black}
\color{black}
\\

\color{black}
One subtle point is the relation between Nambu and Poisson structures. 
It is true that, in some cases, fixing $N-2$ Hamiltonians may induce a Poisson bracket on a reduced leaf. 
However, this does not imply that the original Nambu dynamics is equivalent to an ordinary Hamiltonian system with a single Hamiltonian on the original thermodynamic state space. 
The Nambu bracket keeps the multiple conserved quantities and the associated volume-preserving circulation explicit. 
A standard benchmark illustrating this point is the cyclic Lotka-Volterra system, whose Nambu representation was already described by Frachebourg, Krapivsky, and Ben-Naim\cite{Frachebourg_1996}.\footnote{Frachebourg, Krapivsky, and Ben-Naim showed that the three-species cyclic Lotka-Volterra equations can be written in Nambu form by taking $H_1 = a + b + c$ and $H_2 = abc$ with the canonical Nambu bracket. They also discussed the four-species case with $H_1 = c1 + c2 + c3 + c4$, $H_2 = c1c3$, and $H_3 = c2c4$. We use this system here only as a known benchmark example of Nambu mechanics in cyclic dynamics, not as a new result of the present paper.} We refer to this example only to emphasize that Nambu mechanics naturally keeps multiple conserved quantities explicit in cyclic dynamics. The novelty of the present framework is not this known representation itself, but the use of such Nambu-type volume-preserving circulations as the nondissipative sector of a non-equilibrium thermodynamic decomposition coupled to entropy-gradient dissipation on the same macroscopic state space.
\color{black}
\\

In the GENERIC formalism \cite{_ttinger_1997,_ttinger_2005,Grmela_1997,Grmela_2018},
a structural requirement is imposed such that the entropy function
is a Casimir of the Poisson structure governing the reversible dynamics---that
is, it commutes with the Poisson bracket. As a result, entropy remains
invariant under the reversible part of the dynamics, and its time
evolution is determined solely by the irreversible part, typically
formulated as a gradient flow governed by a friction matrix. This
guarantees compatibility with the second law of thermodynamics, but
at the same time imposes a constraint: entropy must always be degenerate
with respect to the Poisson structure.
\\

In contrast, Nambu Non-equilibrium Thermodynamics adopts a framework
in which the reversible dynamics are governed by a Nambu bracket involving
multiple Hamiltonians, while the irreversible dynamics are driven
by the gradient of the entropy. Within this framework, there is no
structural requirement for entropy to be a Casimir of the Nambu bracket.
In fact, especially for open systems or systems far from equilibrium,
it is often more natural---and more consistent with observed phenomena---not
to impose such a condition. For instance, in systems exhibiting oscillatory
behavior, entropy must be allowed not only to increase via production
but also to decrease due to fluxes out of the system. In such contexts,
it is generally permissible for the reversible part to affect the
entropy, i.e., $\partial_{t}^{(H)}S\neq0$, which marks a clear departure
from the GENERIC framework. Specifically, this structural flexibility
allows for the description of oscillatory non-equilibrium systems
such as the BZ reaction and the Hindmarsh--Rose model, where entropy
can undergo periodic increases and decreases. A detailed discussion
of these models is provided in \cite{NNET-3}.
\\

Thus, while GENERIC represents a constraint-based approach that ensures
thermodynamic consistency via degeneracy conditions imposed on the
Poisson structure, Nambu Non-equilibrium Thermodynamics offers an
extension-based approach. It preserves structural flexibility and
allows for a unified description of competing reversible and dissipative
dynamics.
\\

\color{black}
The above comparison has been made mainly at the level of expressive scope, treating GENERIC in a broad sense as a framework combining a reversible geometric structure with dissipation. In this broad sense, both GENERIC and NNET may be viewed as attempts to unify reversible and irreversible dynamics. However, once one enters a narrower and more structural comparison, the two frameworks differ more fundamentally in the architecture of their state spaces and in the way the reversible structure is introduced. For example, in contact-geometric reformulations of GENERIC such as Grmela et. al.\cite{Grmela_2014}\cite{Esen_2022}\cite{Esen_2022_2}, one works on an extended space involving not only the basic variables but also their conjugate variables, and the reversible/irreversible structures are formulated on that enlarged space. By contrast, NNET does not introduce the Nambu structure by doubling the degrees of freedom and then equipping the enlarged space with an antisymmetric structure. Rather, it starts directly from the given macroscopic thermodynamic state space itself—whether its dimension is even or odd—and constructs the non-dissipative part as a Nambu structure and the dissipative part as an entropy-gradient flow, as a decomposition of the velocity field on that same space. In this sense as well, NNET is not a trivial variant of GENERIC, but a genuinely different geometric formulation of reversible-irreversible coupling.
\color{black}
\\

\color{black}
We also note the relation to metriplectic formulations such as Morrison's bracket formulation for irreversible classical fields\cite{Morrison_1984}. 
There, the dynamics is formulated in terms of generalized Poisson brackets together with symmetric brackets, typically for phase-space densities. 
NNET shares with such approaches the idea of combining nondissipative and dissipative contributions, but differs in that the nondissipative sector is generated by a Nambu bracket directly on the macroscopic thermodynamic state space.
\color{black}

\section{Demonstration of the Formulation Using a Triangular Reaction System}

\color{black}

In this section, we present a concrete formulation of Nambu Non-equilibrium
Thermodynamics (NNET) using a triangular reaction system as an illustrative
example. The triangular reaction has historical significance, since it served
as a foundational model in Onsager's construction of non-equilibrium
thermodynamics, particularly in the near-equilibrium regime\cite{Onsager_1931}.
\\

We first describe the triangular reaction from the viewpoint of chemical
kinetics and recall how Onsager's formulation is recovered under the two
standard assumptions of detailed balance and perturbative deviation from
equilibrium. We then go beyond this limit by introducing a general nonlinear
response expansion of the reaction dynamics.
\\

\color{black}
The purpose of this example is not to rederive standard chemical kinetics or to claim that the dissipative description of chemical reactions is new\footnote{For a standard discussion, see \cite{degroot2013non}.}. Rather, the triangular reaction is used as a minimal example showing how cyclic components hidden beyond the assumptions of detailed balance and linear response can be separated into a Nambu-type circulation and an entropy-gradient contribution.
\color{black}
\\

\color{black}
In this example, the Helmholtz decomposition is used only as a constructive tool. It is not an additional axiom of NNET. Given a velocity field, it provides a way to separate the incompressible component, which is then represented locally by Nambu Hamiltonians, from the compressible gradient component generated by the entropy potential\footnote{Here the Helmholtz decomposition is used locally, in a regular coordinate chart and under the usual regularity assumptions on the velocity field.}. This constructive aspect is used explicitly in Appendix A and is discussed more systematically in \cite{NNET-2}.
\color{black}
\\

It is important to emphasize that this nonlinear expansion is not itself the
Nambu equation. Rather, it provides an organization of the dynamics in powers
of the thermodynamic deviations, and each perturbative contribution is
decomposed, order by order, into a reversible part and an irreversible part
according to the NNET framework introduced in Section 2. In this way, the
circulation-generating sector can be identified systematically and represented
in Nambu form, while the remaining part is described by entropy-gradient-driven
dissipation.
\\

Thus, the purpose of this section is twofold: first, to clarify how the
Onsager limit is embedded in the present framework, and second, to show that
beyond that limit the triangular reaction still admits an order-by-order
NNET decomposition into reversible and irreversible components.
\\

More generally, the reduction of complex nonlinear dynamics to NNET and its extension beyond the perturbative regime are discussed in \cite{NNET-2},
where the role of higher-order mixed tensors and the obstacles to global reduction are analyzed in a broader setting.

\color{black}

\subsection{Triangular Reaction}

The triangular reaction involves three chemical species $X_{1}$,$X_{2}$,$X_{3}$
that undergo the following cyclic reactions\footnote{
\color{black}
The triangular reaction discussed in this section is not treated as an isolated system. Rather, it is considered as an externally driven chemical system characterized by nonzero chemical affinities (or chemical-potential differences). Therefore, the present example should be understood as an open-system illustration of the NNET framework. The discussion of isolated systems given in the previous section is intended to show that NNET can also accommodate the energy-conserving, entropy-producing case under an appropriate geometric choice, but that is logically distinct from the present chemically driven example.
In particular, the entropy function and the chemical potentials appearing here are defined with respect to this driven setting and should not be interpreted as those of an isolated relaxation process.
\color{black}
}:

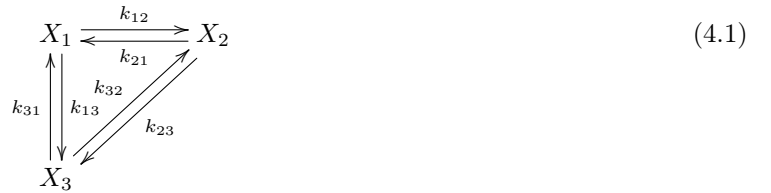
\begin{equation}
\xymatrix@C=4em@R=4em{
X_{1} \ar@<0.5ex>[r]^{k_{12}} \ar@<0.5ex>[d]^{k_{13}} & X_{2} \ar@<0.5ex>[l]^{k_{21}} \ar@<0.5ex>[ld]^{k_{23}} \\
X_{3} \ar@<0.5ex>[u]^{k_{31}} \ar@<0.5ex>[ru]^{k_{32}} &
}
\end{equation}

where $k_{ij}$ denotes the rate constant for the reaction from $X_{i}$
to $X_{j}$.

Let $N_{i}$ be the number of particles of species $X_{i}$, and define
the concentration as $x_{i}=N_{i}/V$, where $V$ is the volume. Assuming
an ideal gas, the equation of state is given by 

\begin{equation}
pV=N\beta^{-1},
\end{equation}
where $p$ is the pressure, $\beta=\frac{1}{k_{B}T}$, and $T$ is
the temperature.

From the Gibbs free energy $G(T,p,N)$, the chemical potential of
the ideal gas satisfies $\beta d\mu=d\log p$. Therefore, the ratio
of concentrations $x_{i}/x_{j}$ can be expressed in terms of the
affinity, defined as the difference in chemical potential $A_{i\to j}\equiv\mu_{i}-\mu_{j}$,
as

\begin{equation}
\frac{x_{i}}{x_{j}}=e^{\beta(A_{i\to j}-A_{(0)i\to j})},
\end{equation}

where $A_{(0)i\to j}$ denotes the equilibrium value of $A_{i\to j}$,
and equilibrium is achieved when $A_{i\to j}=A_{(0)i\to j}$, which
corresponds to $x_{i}=x_{j}$.

Using the above expression together with results from chemical reaction
kinetics, the time evolution of each concentration $x_{i}$ can be
described as follows:
\begin{equation}
\begin{aligned}\frac{dx_{1}}{dt}=-k_{12}\left(1-\frac{k_{21}}{k_{12}}e^{\beta(A_{1\to2}-A_{(0)1\to2})}\right)x_{1}+k_{31}\left(1-\frac{k_{13}}{k_{31}}e^{\beta\left(A_{3\to1}-A_{(0)3\to1}\right)}\right)x_{3},\\
\begin{aligned}\frac{dx_{2}}{dt}=-k_{23}\left(1-\frac{k_{32}}{k_{23}}e^{\beta(A_{2\to3}-A_{(0)2\to3})}\right)x_{2}+k_{12}\left(1-\frac{k_{21}}{k_{12}}e^{\beta\left(A_{1\to2}-A_{(0)1\to2}\right)}\right)x_{1},\\
\begin{aligned}\frac{dx_{3}}{dt}=-k_{31}\left(1-\frac{k_{13}}{k_{31}}e^{\beta(A_{3\to1}-A_{(0)3\to1})}\right)x_{3}+k_{23}\left(1-\frac{k_{32}}{k_{23}}e^{\beta\left(A_{2\to3}-A_{(0)2\to3}\right)}\right)x_{2}.\end{aligned}
\end{aligned}
\end{aligned}
\end{equation}

To recast this system within the conventional framework of Onsager's
non-equilibrium thermodynamics, two assumptions must be introduced.
The first is the principle of detailed balance. Under this assumption,
the following relations are required:
\begin{equation}
-k_{12}x_{1}+k_{21}x_{2}=0,
\end{equation}

\begin{equation}
-k_{23}x_{2}+k_{32}x_{3}=0,
\end{equation}

\begin{equation}
-k_{31}x_{3}+k_{13}x_{1}=0.
\end{equation}

It is important to note that this principle is an assumption deliberately
introduced to enforce relaxation toward equilibrium. There is no compelling
physical reason to impose it in systems that are far from equilibrium.
Therefore, within the axiomatic structure of Nambu Non-equilibrium
Thermodynamics, such an assumption is not adopted as a principle,
but rather considered optional depending on the characteristics of
the system under study. In the present formulation, the detailed balance is assumed only near equilibrium for the Onsager limit.

\LyXZeroWidthSpace{}

The second assumption is the linear approximation for the non-equilibrium
system, under which quantities such as $A_{1\to2}-A_{(0)1\to2}$ are
treated within a linear approximation.

Under these two assumptions, the triangular reaction can be described
by

\begin{equation}
\dot{x}^{i}=L^{ij}\frac{\partial S}{\partial x^{j}},
\end{equation}

where the entropy $S$ is given by

\begin{equation}
S=\Delta\mu_{i}x^{i},
\end{equation}

with $\Delta\mu_{i}$ defined as the deviation of the chemical potential
$\mu_{i}$ from its equilibrium value $\mu_{i}^{(0)}$:

\begin{equation}
\Delta\mu_{i}\equiv\mu_{i}-\mu_{i}^{(0)}.
\end{equation}

The transport coefficients $L^{ij}$ take the following form:

\begin{equation}
L^{11}=\beta\left(k_{12}x_{1}+k_{13}x_{3}\right),
\end{equation}

\begin{equation}
L^{22}=\beta\left(k_{23}x_{2}+k_{21}x_{1}\right),
\end{equation}

\begin{equation}
L^{33}=\beta\left(k_{31}x_{3}+k_{32}x_{2}\right),
\end{equation}

\begin{equation}
L^{12}=L^{21}=-\beta k_{21}x_{1},
\end{equation}

\begin{equation}
L^{23}=L^{32}=-\beta k_{32}x_{2},
\end{equation}

\begin{equation}
L^{31}=L^{13}=-\beta k_{13}x_{3}.
\end{equation}

In Nambu Non-equilibrium Thermodynamics, a more general description
of non-equilibrium systems is possible without assuming either detailed
balance or linearity. To illustrate this point concretely, consider
an expansion in terms of the affinity $A-A_{(0)}$ , and express the
coefficients of the polynomial expansion in $\Delta\mu_{i}$ using
generalized transport tensors $L$.
\\

\color{black}
To extend the triangular reaction beyond the Onsager limit, we now introduce
a general nonlinear response expansion in the thermodynamic deviations
$\Delta \mu_i$. At this stage, the expansion should not be identified with the
Nambu structure itself. Rather, it serves as a response-theoretic decomposition
of the full velocity field into contributions of different perturbative orders.
The role of NNET is to decompose each such contribution, order by order, into
a reversible sector and an irreversible sector. The former is represented, when
possible, in Nambu form, while the latter is described by an entropy-gradient
term.
\color{black}
\\

Then the time evolution of $x^{i}$
can be written as:

\begin{equation}
\frac{dx^{i}}{dt}=\sum_{n=0}^{\infty}\sum_{i_{1},\dots,i_{n}=1}^{3}L_{i,i_{1},\dots,i_{n}}\Delta\mu_{i_{1}}\dots\Delta\mu_{i_{n}}.
\end{equation}

Here, $L_{i,i_{1},\dots,i_{n}}$ is symmetric with respect to the
indices $i_{1},\dots,i_{n}$.
\\

\color{black}
At this stage, the higher-order response tensor itself is not yet the Nambu structure. Rather, it is an intermediate object from which the reversible and irreversible sectors are separated. The symmetric part contributes to the entropy-gradient-driven dissipative flow, while the antisymmetric part, extracted with respect to the distinguished index structure, is reorganized into the Nambu-type nondissipative circulation.
\color{black}
\\

\color{black}
We now organize this expansion by perturbative order and analyze each
contribution separately from the viewpoint of the NNET decomposition.
\color{black}
\\

Expanding the above expression order by order in $\Delta\mu$, we
obtain:

\begin{equation}
\frac{dx^{i}}{dt}=v_{i}^{(0)}+v_{i}^{(1)}+v_{i}^{(2)}\dots,
\end{equation}

where
\begin{equation}
\begin{aligned}v_{i}^{(0)}= & L_{i},\\
v_{i}^{(1)}= & \sum_{i_{1}=1}^{3}L_{i,i_{1}}\Delta\mu_{i_{1}},\\
v_{i}^{(2)}= & \sum_{i_{1},i_{2}=1}^{3}L_{i,i_{1}i_{2}}\Delta\mu_{i_{1}}\Delta\mu_{i_{2}},\\
\vdots
\end{aligned}
\end{equation}

Although in principle higher-order terms can be discussed, we focus
here on terms up to second order.
\\

This is because the terms $v^{(0)}$ and $v^{(2)},v^{(3)},\dots$vanish
under the two assumptions---detailed balance and linear approximation---of
conventional Onsager non-equilibrium thermodynamics.
\\

First, consider the term $L_{i,i_{1}}$, which coincides exactly with
the transport coefficient $L^{ij}$ that appears in Onsager's formulation
of non-equilibrium thermodynamics.
\\

\color{black}
At first order, the contribution reduces to the Onsager transport term,
which in the near-equilibrium regime is purely dissipative in the present
example.
\color{black}
\\

\color{black}
However, for higher-order terms ($n \ge 2$), the generalized tensor $L_{i,i_1,\dots,i_n}$ is symmetric only with respect to the indices $i_1,\dots,i_n$. It generally contains antisymmetric components with respect to the first index $i$ (as explicitly shown later in Eq. (4.30)). In the spirit of NNET, it is precisely these antisymmetric components that are systematically extracted to construct the reversible Nambu bracket (circulation), while the symmetric components are used to construct the irreversible entropy gradient (dissipation).
\color{black}
\\

Next, consider the term $v^{(0)}$, which takes the form:

\begin{equation}
L^{1}=-(k_{12}-k_{21})x_{1}+(k_{31}-k_{13})x_{3},
\end{equation}

\begin{equation}
L^{2}=-(k_{23}-k_{32})x_{2}+(k_{12}-k_{21})x_{1},
\end{equation}

\begin{equation}
L^{3}=-(k_{31}-k_{13})x_{3}+(k_{23}-k_{32})x_{2}.
\end{equation}

These terms are precisely the ones that vanish under the assumption
of detailed balance.

\color{black}
We now analyze these terms order by order. In the spirit of NNET, each
contribution is separated into a reversible part, responsible for circulation,
and an irreversible part, generated by an entropy potential.

We begin with the zeroth-order contribution $v_i^{(0)}$, which survives away
from detailed balance. \color{black}Within the NNET framework, this term admits the following decomposition
into a reversible contribution and an irreversible contribution:\color{black}
\color{black}

\begin{equation}
v_{i}^{(0)}=\sum_{j,k=1}^{3}\epsilon^{ijk}\frac{\partial H_{1}^{(0)}}{\partial x^{j}}\frac{\partial H_{2}^{(0)}}{\partial x^{k}}+\frac{\partial S^{(0)}}{\partial x^{i}}, \label{v0}
\end{equation}

with the components given by:

\begin{equation}
H_{1}^{(0)}=\frac{1}{2}\left(k_{12}-k_{21}\right)x_{3}+\frac{1}{2}\left(k_{31}-k_{13}\right)x_{2}+\frac{1}{2}\left(k_{23}-k_{32}\right)x_{1},
\end{equation}

\begin{equation}
H_{2}^{(0)}=\frac{1}{2}\left(x_{1}^{2}+x_{2}^{2}+x_{3}^{2}\right),
\end{equation}

\begin{equation}
S^{(0)}=-\frac{1}{2}\left((k_{12}-k_{21})(x_{1}+x_{2})x_{1}+(k_{23}-k_{32})(x_{2}+x_{3})x_{2}++(k_{31}-k_{13})(x_{3}+x_{1})x_{3}\right).
\end{equation}

\color{black}
The first term in Eq. (\ref{v0}) is the Nambu, or circulation-generating, component of the zeroth-order reaction velocity field, whereas the second term is the entropy-gradient component. Thus, even at this order, the reaction dynamics is not treated as purely dissipative once detailed balance is relaxed.
\color{black}
\\

Here, $S^{(0)}$represents a dissipative structure that arises due
to asymmetries in the reaction rates. Likewise, $H_{1}^{(0)}$vanishes
when the rate asymmetries are eliminated. In contrast, $H_{2}^{(0)}$is
a geometric conserved quantity independent of such asymmetries, corresponding
to the \textquotedbl squared radius\textquotedbl{} in the reaction
space and reflecting the underlying cyclic structure.

\color{black}
Next, we turn to the second-order contribution $v_i^{(2)}$. Our purpose here
is to separate the circulation-generating part from the dissipative part,
again in accordance with the NNET decomposition. For this reason, it is useful
to distinguish the symmetric and antisymmetric structures contained in the
second-order transport coefficients. Its components are explicitly given by:
\color{black}
\begin{equation}
\begin{aligned}L_{1,11} & =\frac{1}{2}\beta^{2}\left(k_{21}x_{1}-k_{13}x_{3}\right),\\
L_{2,22} & =\frac{1}{2}\beta^{2}\left(k_{32}x_{2}-k_{21}x_{1}\right),\\
L_{3,33} & =\frac{1}{2}\beta^{2}\left(k_{13}x_{3}-k_{32}x_{2}\right),\\
L_{1,22} & =-L_{2,11}=\frac{1}{2}\beta^{2}k_{21}x_{1},\\
L_{2,33} & =-L_{3,22}=\frac{1}{2}\beta^{2}k_{32}x_{2},\\
L_{3,11} & =-L_{1,33}=\frac{1}{2}\beta^{2}k_{13}x_{3},\\
L_{1,12} & =-L_{2,21}=-\frac{1}{2}\beta^{2}k_{21}x_{1},\\
L_{2,23} & =-L_{3,32}=-\frac{1}{2}\beta^{2}k_{32}x_{2},\\
L_{3,31} & =-L_{1,13}=-\frac{1}{2}\beta^{2}k_{13}x_{3}.
\end{aligned}
\end{equation}

\color{black}
For the present triangular reaction, these coefficients satisfy the relations
\color{black}

\begin{equation}
L_{i,i_{1}i_{2}}=L_{i_{1},ii_{2}}=L_{i_{2},ii_{1}}.
\end{equation}

In addition, the coefficients exhibit the following antisymmetric
properties:

\begin{equation}
L_{i,jj}=-L_{j,ii},\ L_{i,ij}=-L_{j,ji}.
\end{equation}

To facilitate the decomposition, we introduce the following symmetric
tensor:
\begin{equation}
\tilde{A}_{(ij)}\equiv\frac{1}{2}\left(\Delta\mu_{i}\Delta\mu_{j}+\Delta\mu_{j}\Delta\mu_{i}\right).
\end{equation}
Although $\Delta\mu_{i}\Delta\mu_{j}$ is symmetric under index exchange,
we introduce the symmetrized notation $\tilde{A}_{(ij)}$ to emphasize
its role as a symmetric second-rank tensor in the decomposition of
higher-order transport terms. This allows us to clearly separate symmetric
and antisymmetric components in the tensorial structure of the second-order
contributions.

We define the following antisymmetric tensor:
\begin{equation}
B_{i,j}\equiv L_{i,jj}
\end{equation}

Using these, the second-order term $v_{i}^{(2)}$ can be decomposed
into its antisymmetric and symmetric parts as follows:
\begin{equation}
v_{i}^{(2)}=\sum_{j=1}^{3}B_{i,j}\tilde{A}_{(jj)}+\sum_{j=1}^{3}L_{i,ij}\tilde{A}_{(ij)}.
\end{equation}

\color{black}In the framework of Nambu Non-equilibrium Thermodynamics, this second-order
term admits a corresponding decomposition. One convenient choice is as follows.\color{black}

First, we choose $H_{1}^{(2)}$
such that its gradient yields $\tilde{A}_{(jj)}$. Accordingly, we
define
\begin{equation}
H_{1}^{(2)}=\sum_{i=1}^{3}\tilde{A}_{(ii)}x^{i}. \label{H1_2nd}
\end{equation}

Next, we determine $H_{2}^{(2)}$ from $B_{i,j}$ by requiring that
it satisfies

\begin{equation}
\epsilon^{ijk}\frac{\partial H_{2}^{(2)}}{\partial x^{k}}=B_{i,j},
\end{equation}

which yields
\begin{equation}
H_{2}^{(2)}=\beta^{2}\left(k_{21}x^{1}x^{3}+k_{32}x^{2}x^{1}+k_{13}x^{3}x^{2}\right). \label{H2_2nd}
\end{equation}

Furthermore, the entropy component $S^{(2)}$ is determined by the
symmetric second-order tensor through the condition:
\begin{equation}
\sum_{j=1}^{3}L_{i,ij}\tilde{A}_{i,j}=\sum_{j=1}^{3}L_{i,ij}\frac{\partial^{2}S^{(2)}}{\partial x^{i}\partial x^{j}}.\label{eq:LA=00003DLS2}
\end{equation}
Solving this yields:

\begin{equation}
S^{(2)}=\sum_{i,j=1}^{3}\tilde{A}_{(ij)}x^{i}x^{j}.\label{eq:S2}
\end{equation}

Here, $\tilde{A}_{i,j}$ in Eq.(\ref{eq:LA=00003DLS2}) denotes the
components of a symmetric tensor arising from the second derivatives
of $S^{(2)}$, which is explicitly constructed from the symmetrized
expression$\tilde{A}_{(ij)}$ in Eq. (\ref{eq:S2}).
\\

\color{black}
Equations (\ref{H1_2nd})-(\ref{H2_2nd}) explicitly identify the Nambu component of the second-order contribution. In contrast, Eqs. (\ref{eq:LA=00003DLS2})-(\ref{eq:S2}) determine the entropy-gradient component. This separation is the concrete realization of the NNET decomposition for the triangular reaction beyond the Onsager limit.
\color{black}
\\

Here again, it is important to note that both $S^{(2)}$ and $H_{1}^{(2)}$vanish
in the vicinity of equilibrium---i.e., when the assumptions of detailed
balance and linearity (Assumption 2) hold. In contrast, $H_{2}^{(2)}$
remains nonzero and represents a geometric conserved quantity that
is independent of such assumptions. This distinction becomes apparent
only through the formalism of Nambu Non-equilibrium Thermodynamics\footnote{The generalization of this framework to nonlinear thermodynamic regimes
far from equilibrium will be explored in greater depth in subsequent work\cite{NNET-2}.}.

This example highlights the significance of the broad descriptive
power of Nambu Non-equilibrium Thermodynamics, which does not rely
on the assumptions of detailed balance or linear response that are
central to Onsager's theory. It also underscores the physical meaning
of each Hamiltonian in this extended formalism.

\section{Conclusion and Discussion}

In this study, we have introduced an axiomatic formulation of non-equilibrium
thermodynamics based on the Nambu bracket, termed Nambu Non-equilibrium
Thermodynamics (NNET). Through a series of lemmas derived from the
proposed axioms, we have demonstrated that this framework is capable
of describing far-from-equilibrium systems in which entropy may decrease
due to periodic dynamics---phenomena that may be difficult to capture
using Prigogine's General Evolution Criterion (GEC) or Grmela--$\ddot{\mathrm{O}}$ttinger's
GENERIC formalism. Our system is also open; hence there is no contradiction with the open-system nature of GEC.
\\

As a concrete example, we analyzed the triangular reaction system
and showed that NNET allows for a description that goes beyond the
conventional assumptions of detailed balance and linear approximation.
In doing so, we revealed the existence of geometric conserved quantities
that are otherwise obscured under those assumptions.
\\

\color{black}
Appendix A further shows that, in the representative extended-variable GENERIC-style formulation considered there, the same triangular reaction is formulated on an enlarged state space with flux variables, whereas NNET decomposes the dynamics directly on the original macroscopic state space. This indicates that the difference between the two frameworks is not merely technical but structural.
\color{black}
\\

A systematic treatment of more general nonlinear phenomena will be
developed in  \cite{NNET-2}. Applications to systems exhibiting
periodic oscillations or spike-like behavior---such as the BZ reaction
and the Hindmarsh--Rose model---will be explored in \cite{NNET-3}.
\\

\color{black}
Another important direction is the extension of NNET to field-theoretic and hydrodynamic systems. The present paper is restricted to finite-dimensional macroscopic state spaces, and we do not claim to have completed such an extension here. Nevertheless, because advective and incompressible components in fluid dynamics are naturally related to Nambu-type structures, it would be interesting to derive hydrodynamic NNET equations from local-equilibrium assumptions and to clarify their relation to kinetic theory and continuum thermodynamics.
\color{black}
\\

\color{black}
A statistical description including thermal fluctuations and stochastic
processes is fundamental to a microscopic understanding of Nambu
Non-equilibrium Thermodynamics. Although a systematic treatment of this
aspect is beyond the scope of the present paper, essential elements have
already been developed in our earlier work \cite{https://doi.org/10.48550/arxiv.2209.08469},
particularly in connection with the OMH framework \cite{Onsager_1953,Hashitsume_1952}
and Zwanzig's model \cite{Zwanzig_1973}. It is also important to compare the
present formulation with the large-deviation and generalized-gradient
viewpoints developed by Mielke, Peletier, and Renger \cite{Mielke_2014}, as well
as with the subsequent decomposition of dissipative and Hamiltonian
contributions discussed by Renger and Sharma \cite{Renger_2023}. However, the
present framework is not intended to replace formulations based on
dissipation potentials or large-deviation principles. Rather, our standpoint
is complementary: the aim of NNET is to make the Nambu-type
nondissipative sector explicit within a unified decomposition of the velocity
field. In this respect, the decomposition of dissipative and Hamiltonian
effects in large-deviation-based approaches is conceptually related to the
present one, although the geometric origin of the reversible structure is
different.
\color{black}
\\

One important question that arises concerns the origin of the entropy
function introduced in this framework: does it correspond to the physical
entropy as traditionally understood? In fact, the entropy discussed
here serves as a potential function responsible for generating the
dissipative term in the dynamical system. It aligns with the conventional
notion of entropy in non-equilibrium thermodynamics only under the
two assumptions imposed by Onsager's theory. This suggests that entropy,
when far from equilibrium, may carry multiple meanings.
\\

For instance, in traditional thermodynamics, as illustrated by textbook
descriptions of piston systems, entropy is typically conserved during
quasi-static reversible processes. Within the NNET framework, however,
entropy can also be understood as a measure of deviation from such
reversible trajectories. This leads to a pluralistic picture of entropy:
one associated with conserved quantities in quasi-static limits, and
another representing dissipative departures from them. A more detailed
analysis of this perspective will be presented in future work \cite{NNET-Piston}.
\\

As Schr\"odinger once remarked, 'Life feeds on negative entropy.'.
We hope that the discussion of Nambu Non-equilibrium Thermodynamics
offers new insight into the description of complex systems such as
the ocean or biological structures---systems that have traditionally
resisted rigorous analysis within the existing thermodynamic frameworks.

\section*{Acknowledgments}

We would like to thank Toshio Fukumi for his advice on non-linear
response theory. We are also grateful to Shiro Komata for carefully
reading the manuscript and providing valuable comments.

\color{black}
\appendix

\section{Triangular Reaction Beyond the Perturbative Treatment}

In the main text, the triangular reaction was treated perturbatively in order to facilitate comparison with Onsager theory. However, NNET can also treat the same system non-perturbatively. This appendix summarizes that non-perturbative treatment; for further details, see Part 2.

For comparison, we also sketch a representative GENERIC-style extended-variable formulation of the same triangular reaction, in order to highlight the structural difference between the NNET and GENERIC viewpoints.

\subsection{Triangular Reaction in NNET}

As a concrete example, we consider a triangular chemical reaction and explicitly construct the functions $H_{1}$, $H_{2}$, and $S$ by applying a Helmholtz decomposition to the corresponding velocity field.

The triangular reaction is the cyclic chemical process
\[
X_{1}\to X_{2}\to X_{3}\to X_{1},
\]
which may be represented schematically as
\begin{equation}
\xymatrix@C=4em@R=4em{
X_{1} \ar@<0.5ex>[r]^{k_{12}} \ar@<0.5ex>[d]^{k_{13}} & X_{2} \ar@<0.5ex>[l]^{k_{21}} \ar@<0.5ex>[ld]^{k_{23}} \\
X_{3} \ar@<0.5ex>[u]^{k_{31}} \ar@<0.5ex>[ru]^{k_{32}} &
}
\end{equation}

The corresponding rate equations are
\begin{equation}
\frac{dx_{1}}{dt}=-\tilde{k}_{12}x_{1}+\tilde{k}_{31}x_{3},
\end{equation}
\begin{equation}
\frac{dx_{2}}{dt}=-\tilde{k}_{23}x_{2}+\tilde{k}_{12}x_{1},
\end{equation}
\begin{equation}
\frac{dx_{3}}{dt}=-\tilde{k}_{31}x_{3}+\tilde{k}_{23}x_{2}.
\end{equation}

Here the coefficients $\tilde{k}_{ij}$ are defined in terms of the bare reaction rates $k_{ij}$ and the thermodynamic affinities $A_{i\to j}$ by
\begin{equation}
\tilde{k}_{ij}\equiv k_{ij}\left(1-\frac{k_{ji}}{k_{ij}}e^{\beta(A_{i\to j}-A_{i\to j}^{(0)})}\right),
\end{equation}
where $\beta$ denotes the inverse temperature, $A_{i\to j}$ is the affinity between species $X_i$ and $X_j$, and $A_{i\to j}^{(0)}$ denotes its equilibrium value.

We now apply a Helmholtz decomposition to separate the flow into compressible and incompressible parts, and then construct the corresponding Hamiltonians and entropy scalar. More precisely, we first decompose the three-dimensional vector field into a solenoidal part and a gradient part, and then represent the solenoidal part in Nambu form as
\begin{equation}
B=\nabla H_{1}\times\nabla H_{2}.
\end{equation}

\subsubsection*{Helmholtz decomposition}

In three dimensions, the Helmholtz decomposition allows any vector field $V^{i}=\dot{x}^{i}$ to be written as
\begin{equation}
\dot{x}^{i}=\phi^{i}+B^{i}+\frac{\partial S}{\partial x^{i}},
\end{equation}
where $\phi^{i}$ satisfies $\Delta\phi^{i}=0$, $B^{i}$ is constructed from an antisymmetric tensor $B_{jk}$ as $B^{i}=\epsilon^{ijk}B_{jk}$, and $\partial S/\partial x^{i}$ is derived from a scalar potential $S$.

Let us determine $\phi$, $B^{i}$, and $S$ for the triangular reaction. For simplicity, we set $\phi=0$. We then determine $S$ first, and obtain $B^{i}$ from
\[
B^{i}=V^{i}-\frac{\partial S}{\partial x^{i}}.
\]

We choose a particular scalar potential $S$ satisfying both
\begin{equation}
\nabla\cdot V=\Delta S=-k_{\Sigma},
\end{equation}
and
\begin{equation}
\nabla H_{1}\cdot\nabla S=0,
\end{equation}
with
\begin{equation}
k_{\Sigma}\equiv \tilde{k}_{12}+\tilde{k}_{23}+\tilde{k}_{31}.
\end{equation}

Next, we observe that the triangular reaction has the conserved quantity
\begin{equation}
x_{1}+x_{2}+x_{3},
\end{equation}
and therefore set
\begin{equation}
H_{1}\equiv x_{1}+x_{2}+x_{3}.
\end{equation}
Since $H_{1}$ is conserved, it satisfies
\begin{equation}
\nabla H_{1}\cdot V=0.
\end{equation}

A particular choice of $S$ satisfying the above conditions is
\begin{equation}
\nabla S=\frac{k_{\Sigma}}{2}\left(-x_{1}+x_{2},\, -x_{2}+x_{1},\, 0\right),
\end{equation}
which yields
\begin{equation}
S=-\frac{k_{\Sigma}}{4}(x_{1}-x_{2})^{2}.
\end{equation}

The remaining part $B^{i}$ is then given by
\begin{equation}
B^{i}=
\begin{pmatrix}
-\left(\tilde{k}_{12}-\frac{k_{\Sigma}}{2}\right) & -\frac{k_{\Sigma}}{2} & \tilde{k}_{31}\\
\left(\tilde{k}_{12}-\frac{k_{\Sigma}}{2}\right) & -\left(\tilde{k}_{23}-\frac{k_{\Sigma}}{2}\right) & 0\\
0 & \tilde{k}_{23} & -\tilde{k}_{31}
\end{pmatrix}
\begin{pmatrix}
x_{1}\\
x_{2}\\
x_{3}
\end{pmatrix}.
\end{equation}

\subsubsection*{Darboux part}

The remaining divergence-free part is represented by a linear vector field $B=Mx$, and we seek $H_{2}$ such that
\[
Mx=\nabla H_{1}\times\nabla H_{2}.
\]

Assuming a local canonical representation of the divergence-free part, we determine $H_{2}$ from
\begin{equation}
B=\nabla H_{1}\times\nabla H_{2}
=
\begin{pmatrix}
1\\
1\\
1
\end{pmatrix}
\times \nabla H_{2}.
\end{equation}

Introducing a matrix $M$, we require
\begin{equation}
\begin{pmatrix}
1\\
1\\
1
\end{pmatrix}
\times \nabla H_{2}
=
M
\begin{pmatrix}
x_{1}\\
x_{2}\\
x_{3}
\end{pmatrix},
\end{equation}
which yields
\begin{equation}
M=
\begin{pmatrix}
a & b & c\\
a & b+\tilde{k}_{23} & c-\tilde{k}_{31}\\
a-\tilde{k}_{12}+\frac{k_{\Sigma}}{2} & b+\tilde{k}_{23}-\frac{k_{\Sigma}}{2} & c
\end{pmatrix}.
\end{equation}

Here $a$, $b$, and $c$ are not independent, but satisfy
\begin{equation}
a=b=c+\frac{\tilde{k}_{12}-\tilde{k}_{23}-\tilde{k}_{31}}{2}.
\end{equation}

Choosing
\begin{equation}
c=-\frac{\tilde{k}_{12}-\tilde{k}_{23}-\tilde{k}_{31}}{2},
\end{equation}
so that $a=b=0$, we obtain
\begin{equation}
H_{2}
=
\frac{1}{2}\left(c-\tilde{k}_{12}+\frac{k_{\Sigma}}{2}\right)x_{3}x_{1}
+\frac{1}{2}\tilde{k}_{23}x_{2}^{2}
+\frac{1}{2}\left(c-\tilde{k}_{31}+\tilde{k}_{23}-\frac{k_{\Sigma}}{2}\right)x_{2}x_{3}.
\end{equation}

Thus, for the triangular reaction we obtain the set $(H_{1},H_{2},S)$:
\begin{equation}
H_{1}=x_{1}+x_{2}+x_{3},
\end{equation}
\begin{equation}
H_{2}
=
\frac{1}{2}\left(c-\tilde{k}_{12}+\frac{k_{\Sigma}}{2}\right)x_{3}x_{1}
+\frac{1}{2}\tilde{k}_{23}x_{2}^{2}
+\frac{1}{2}\left(c-\tilde{k}_{31}+\tilde{k}_{23}-\frac{k_{\Sigma}}{2}\right)x_{2}x_{3},
\end{equation}
\begin{equation}
S=-\frac{k_{\Sigma}}{4}(x_{1}-x_{2})^{2},
\end{equation}
with
\begin{equation}
c=-\frac{\tilde{k}_{12}-\tilde{k}_{23}-\tilde{k}_{31}}{2}.
\end{equation}

\paragraph{Remark.}
It is also possible not to choose a conserved quantity as $H_{1}$. For example, one may instead take
\begin{equation}
H_{1}=x_{1}^{2}+x_{2}^{2}+x_{3}^{2},
\end{equation}
in which case $H_{2}$ and $S$ take more symmetric forms. Which choice is most appropriate depends on the context.

\subsection{Triangular Reaction in a Representative GENERIC-style Formulation}

For chemical kinetics in a GENERIC-related setting, see, for example,
\cite{Grmela_2012}\cite{Grmela_2021}\cite{Ajji_2023}. 
For comparison with the NNET construction, we now present
a representative extended-variable GENERIC-style formulation
of the same triangular reaction.

\subsubsection*{Reaction channels and stoichiometric structure}

We consider the three reaction channels

\[
r=1:\; X_{1}\to X_{2}, \qquad
r=2:\; X_{2}\to X_{3}, \qquad
r=3:\; X_{3}\to X_{1}.
\]

Let $x_i$ ($i=1,2,3$) denote the concentrations of species $X_i$.
The time evolution of the concentrations is expressed in terms
of the reaction fluxes $J_r$ through the stoichiometric
incidence matrix $\gamma_{ir}$:

\begin{equation}
\dot{x}_i
=
\sum_{r=1}^{3}
\gamma_{ir}\,J_r .
\end{equation}

For the triangular reaction, the matrix $\gamma_{ir}$ is defined as

\begin{equation}
\gamma
=
\begin{pmatrix}
-1 & 0 & 1 \\
\;\,1 & -1 & 0 \\
\;\,0 & 1 & -1
\end{pmatrix}.
\end{equation}

Thus the evolution equations explicitly become

\begin{align}
\dot{x}_1 &= -J_1 + J_3, \\
\dot{x}_2 &= \;\,J_1 - J_2, \\
\dot{x}_3 &= \;\,J_2 - J_3.
\end{align}

\subsubsection*{Affinity variables and dissipation potential}

Next, introduce the thermodynamic affinities

\begin{equation}
a_r
=
\beta
\left(
A_r - A_r^{(0)}
\right),
\end{equation}

where $A_r$ denotes the affinity associated with reaction channel $r$,
$A_r^{(0)}$ its equilibrium value,
and $\beta$ is the inverse temperature.

We then introduce a dissipation potential
$\Xi_{\mathrm{Generic}}(x,a)$ of the form

\begin{equation}
\Xi_{\mathrm{Generic}}
=
\sum_{r=1}^{3}
W_r(x)
\left(
e^{a_r/2}
+
e^{-a_r/2}
-
2
\right).
\end{equation}

A typical choice for the mobility factors is

\begin{equation}
W_1(x)
=
\sqrt{k_{12}k_{21}x_1x_2},
\end{equation}

\begin{equation}
W_2(x)
=
\sqrt{k_{23}k_{32}x_2x_3},
\end{equation}

\begin{equation}
W_3(x)
=
\sqrt{k_{31}k_{13}x_3x_1}.
\end{equation}

The reaction fluxes are generated from the dissipation potential by

\begin{equation}
J_r
=
-
\frac{\partial
\Xi_{\mathrm{Generic}}
}{\partial a_r}.
\end{equation}

Explicitly,

\begin{equation}
J_r
=
-
W_r(x)
\sinh\!\left(\frac{a_r}{2}\right).
\end{equation}

\subsubsection*{Legendre transform and extended variables}

To construct an extended-variable formulation,
we introduce the Legendre transform of
$\Xi_{\mathrm{Generic}}(x,a)$
with respect to $a_r$:

\begin{equation}
\Xi_{\mathrm{Generic}}^{\dagger}(x,J)
=
\sup_{a}
\left[
\sum_{r=1}^{3}
a_r J_r
-
\Xi_{\mathrm{Generic}}(x,a)
\right].
\end{equation}

In this formulation, the reaction fluxes $J_r$
are treated as independent dynamical variables,
and the system is described on the enlarged
state space $(x,J)$.

\subsubsection*{Extended Hamiltonian structure}

On the enlarged state space $(x,J)$,
we introduce an extended Hamiltonian
(or energy-like functional)

\begin{equation}
H_{\mathrm{ext}}(x,J)
=
\Phi(x)
+
\frac{1}{2}
\sum_{r=1}^{3}
J_r^{2}.
\end{equation}

Here:

\begin{itemize}

\item
$\Phi(x)$ denotes a thermodynamic potential
defined on the macroscopic variables $x_i$
(e.g., a free-energy-like function).

\end{itemize}

\subsubsection*{Coupled evolution equations}

The coupled evolution equations
on the enlarged space $(x,J)$
are then written as

\begin{equation}
\dot{x}_i
=
\sum_{r=1}^{3}
\gamma_{ir}
\frac{\partial
H_{\mathrm{ext}}
}{\partial J_r},
\end{equation}

\begin{equation}
\dot{J}_r
=
-
\sum_{i=1}^{3}
\gamma_{ir}
\frac{\partial
H_{\mathrm{ext}}
}{\partial x_i}
-
\frac{\partial
\Xi_{\mathrm{Generic}}^{\dagger}
}{\partial J_r}.
\end{equation}

The first term represents the reversible
antisymmetric coupling between the
concentration variables and the flux variables,
while the second term generates the dissipative
relaxation associated with the Legendre-transformed
dissipation potential.

\subsubsection*{Detailed balance condition}

The above formulation implicitly incorporates the detailed balance structure through the symmetric choice of the mobility factors $W_r(x)$ and the definition of the affinity variables $a_r$ relative to their equilibrium values. In particular, at equilibrium one has $a_r=0$, and therefore the fluxes satisfy $J_r=0$. The symmetric form of $W_r(x)$ with respect to forward and backward reactions ensures compatibility with detailed balance at the level of individual reaction channels.

\subsubsection*{Structural comparison with NNET}

The purpose of this construction is not to provide
a full thermodynamic model of the triangular reaction
within the GENERIC framework,
but rather to highlight the structural difference
between the GENERIC-style extended formulation
and the NNET formulation.

In the GENERIC-style description above,
the reversible--irreversible coupling is constructed
on an enlarged state space $(x,J)$,
in which the reaction fluxes are introduced
as additional dynamical variables.

By contrast,
in the NNET formulation presented in
Section~A.1,
the decomposition into non-dissipative
and dissipative parts is carried out
directly on the original macroscopic
state space $x$ itself,
without introducing additional flux variables.

Thus, even for the same triangular reaction,
the difference between NNET and GENERIC
appears already at the level of the
degrees of freedom and the architecture
of the state space.
\color{black}
\bibliographystyle{unsrturl}
\bibliography{NambuNTD_I_Foundation_en_v4}

@Article{Onsager_1931,
  author    = {Onsager, Lars},
  journal   = {Physical Review},
  title     = {Reciprocal {R}elations in {I}rreversible {P}rocesses. {I}.},
  year      = {1931},
  issn      = {0031-899X},
  month     = feb,
  number    = {4},
  pages     = {405--426},
  volume    = {37},
  doi       = {https://doi.org/10.1103/PhysRev.37.405},
  publisher = {American Physical Society (APS)},
}

@Article{Onsager_1931_2,
  author    = {Onsager, Lars},
  journal   = {Physical Review},
  title     = {Reciprocal {R}elations in {I}rreversible {P}rocesses. {II}.},
  year      = {1931},
  issn      = {0031-899X},
  month     = dec,
  number    = {12},
  pages     = {2265--2279},
  volume    = {38},
  doi       = {https://doi.org/10.1103/PhysRev.38.2265},
  publisher = {American Physical Society (APS)},
}

@Article{Glansdorff_1964,
  author    = {Glansdorff, P. and Prigogine, I.},
  journal   = {Physica},
  title     = {On a {G}eneral {E}volution {C}riterion in {M}acroscopic {P}hysics},
  year      = {1964},
  issn      = {0031-8914},
  month     = feb,
  number    = {2},
  pages     = {351--374},
  volume    = {30},
  doi       = {https://doi.org/10.1016/0031-8914(64)90009-6},
  publisher = {Elsevier BV},
}

@Article{Grmela_1997,
  author    = {Grmela, Miroslav and {\"O}ttinger, Hans Christian},
  journal   = {Physical Review E},
  title     = {Dynamics and {T}hermodynamics of {C}omplex fluids. {I}. {D}evelopment of a general formalism},
  year      = {1997},
  issn      = {1095-3787},
  month     = dec,
  number    = {6},
  pages     = {6620--6632},
  volume    = {56},
  doi       = {https://doi.org/10.1103/PhysRevE.56.6620},
  publisher = {American Physical Society (APS)},
}

@Article{_ttinger_1997,
  author    = {{\"O}ttinger, Hans Christian and Grmela, Miroslav},
  journal   = {Physical Review E},
  title     = {Dynamics and {T}hermodynamics of {C}omplex {F}luids. {II}. {I}llustrations of a general formalism},
  year      = {1997},
  issn      = {1095-3787},
  month     = dec,
  number    = {6},
  pages     = {6633--6655},
  volume    = {56},
  doi       = {https://doi.org/10.1103/PhysRevE.56.6633},
  publisher = {American Physical Society (APS)},
}

@Article{_ttinger_2005,
  author    = {{\"O}ttinger, Hans Christian},
  journal   = {John Wiley \& Sons, Inc},
  title     = {Beyond {E}quilibrium {T}hermodynamics},
  year      = {2005},
  month     = jan,
  doi       = {10.1002/0471727903},
  isbn      = {9780471727903},
  publisher = {Wiley},
}

@Article{Grmela_2018,
  author    = {Grmela, Miroslav},
  journal   = {Journal of Physics Communications},
  title     = {GENERIC {G}uide to {T}he {M}ultiscale {D}ynamics and {T}hermodynamics},
  year      = {2018},
  issn      = {2399-6528},
  month     = mar,
  number    = {3},
  pages     = {032001},
  volume    = {2},
  doi       = {10.1088/2399-6528/aab642},
  publisher = {IOP Publishing},
}

@Article{https://doi.org/10.48550/arxiv.2209.08469,
  author    = {Katagiri, So and Matsuoka, Yoshiki and Sugamoto, Akio},
  title     = {Fluctuating {N}on-linear {N}on-equilibrium {S}ystem in {T}erms of {N}ambu {T}hermodynamics},
  year      = {2022},
  copyright = {arXiv.org perpetual, non-exclusive license},
  doi       = {https://doi.org/10.48550/arXiv.2209.08469},
  publisher = {arXiv},
}

@Article{NNET-2,
  author    = {Katagiri, So and Matsuoka, Yoshiki and Sugamoto, Akio},
  journal   = {Journal of Mathematical Physics},
  title     = {Reduction of complex dynamics in far-from-equilibrium systems: Nambu non-equilibrium thermodynamics},
  year      = {2026},
  issn      = {1089-7658},
  month     = Apr,
  number    = {4},
  volume    = {67},
  doi       = {https://doi.org/10.1063/5.0309729},
  publisher = {AIP Publishing},
}

@Article{NNET-3,
  author    = {Katagiri, So and Matsuoka, Yoshiki and Sugamoto, Akio},
  journal   = {preprint},
  title     = {{A}pplications of {N}ambu {N}on-equilibrium {T}hermodynamics to {S}pecific {P}henomena},
  year      = {2025},
  copyright = {arXiv.org perpetual, non-exclusive license},
  doi       = {https://doi.org/10.48550/arXiv.2509.12641},
  publisher = {arXiv},
}

@Article{NNET-Piston,
  author  = {Katagiri, So},
  journal = {in progress},
  title   = {Nambu {N}on-equilibrium {T}hermodynamics of a {P}iston {S}ystem},
}

@Article{Nambu_1973,
  author    = {Nambu, Yoichiro},
  journal   = {Physical Review D},
  title     = {Generalized {H}amiltonian {D}ynamics},
  year      = {1973},
  issn      = {0556-2821},
  month     = apr,
  number    = {8},
  pages     = {2405--2412},
  volume    = {7},
  doi       = {https://doi.org/10.1103/PhysRevD.7.2405},
  publisher = {American Physical Society (APS)},
}

@Article{Kondepudi_2014,
  author    = {Kondepudi, Dilip and Prigogine, Ilya},
  journal   = {John Wiley \& Sons, Inc},
  title     = {Modern {T}hermodynamics: {F}rom {H}eat {E}ngines to {D}issipative Structures},
  year      = {2014},
  month     = nov,
  doi       = {10.1002/9781118698723},
  isbn      = {9781118698723},
  publisher = {Wiley},
}

@Article{Zwanzig_1973,
  author    = {Zwanzig, Robert},
  journal   = {Journal of Statistical Physics},
  title     = {Nonlinear {G}eneralized {L}angevin {E}quations},
  year      = {1973},
  issn      = {1572-9613},
  month     = nov,
  number    = {3},
  pages     = {215--220},
  volume    = {9},
  doi       = {https://doi.org/10.1007/BF01008729},
  publisher = {Springer Science and Business Media LLC},
}

@Article{Onsager_1953,
  author    = {Onsager, L. and Machlup, S.},
  journal   = {Physical Review},
  title     = {Fluctuations and {I}rreversible {P}rocesses},
  year      = {1953},
  issn      = {0031-899X},
  month     = sep,
  number    = {6},
  pages     = {1505--1512},
  volume    = {91},
  doi       = {https://doi.org/10.1103/PhysRev.91.1505},
  publisher = {American Physical Society (APS)},
}

@Article{Hashitsume_1952,
  author    = {Hashitsume, N.},
  journal   = {Progress of Theoretical Physics},
  title     = {A {S}tatistical {T}heory of {L}inear {D}issipative {S}ystems},
  year      = {1952},
  issn      = {1347-4081},
  month     = oct,
  number    = {4},
  pages     = {461--478},
  volume    = {8},
  doi       = {https://doi.org/10.1143/ptp/8.4.461},
  publisher = {Oxford University Press (OUP)},
}

@Article{Frachebourg_1996,
  author    = {Frachebourg, L. and Krapivsky, P. L. and Ben-Naim, E.},
  journal   = {Physical Review E},
  title     = {Spatial {O}rganization in {C}yclic {L}otka-{V}olterra {S}ystems},
  year      = {1996},
  issn      = {1095-3787},
  month     = dec,
  number    = {6},
  pages     = {6186--6200},
  volume    = {54},
  doi       = {https://doi.org/10.1103/PhysRevE.54.6186},
  publisher = {American Physical Society (APS)},
}

@Article{Esen_2022,
  author    = {Esen, Oğul and Grmela, Miroslav and Pavelka, Michal},
  journal   = {Journal of Mathematical Physics},
  title     = {On the role of geometry in statistical mechanics and thermodynamics. I. Geometric perspective},
  year      = {2022},
  issn      = {1089-7658},
  month     = dec,
  number    = {12},
  volume    = {63},
  doi       = {https://doi.org/10.1063/5.0099923},
  publisher = {AIP Publishing},
}

@Article{Esen_2022_2,
  author    = {Esen, Oğul and Grmela, Miroslav and Pavelka, Michal},
  journal   = {Journal of Mathematical Physics},
  title     = {On the role of geometry in statistical mechanics and thermodynamics. II. Thermodynamic perspective},
  year      = {2022},
  issn      = {1089-7658},
  month     = dec,
  number    = {12},
  volume    = {63},
  doi       = {https://doi.org/10.1063/5.0099930},
  publisher = {AIP Publishing},
}

@Article{Grmela_2014,
  author    = {Grmela, Miroslav},
  journal   = {Entropy},
  title     = {Contact Geometry of Mesoscopic Thermodynamics and Dynamics},
  year      = {2014},
  issn      = {1099-4300},
  month     = mar,
  number    = {3},
  pages     = {1652--1686},
  volume    = {16},
  doi       = {doi:10.3390/e16031652},
  publisher = {MDPI AG},
}

@Article{Ajji_2023,
  author    = {Ajji, Abdellah and Chaouki, Jamal and Esen, Oğul and Grmela, Miroslav and Klika, Václav and Pavelka, Michal},
  journal   = {Physica D: Nonlinear Phenomena},
  title     = {On geometry of multiscale mass action law and its fluctuations},
  year      = {2023},
  issn      = {0167-2789},
  month     = mar,
  pages     = {133642},
  volume    = {445},
  doi       = {https://doi.org/10.1016/j.physd.2022.133642},
  publisher = {Elsevier BV},
}

@Article{Grmela_2012,
  author    = {Grmela, Miroslav},
  journal   = {Physica D: Nonlinear Phenomena},
  title     = {Fluctuations in extended mass-action-law dynamics},
  year      = {2012},
  issn      = {0167-2789},
  month     = may,
  number    = {10},
  pages     = {976--986},
  volume    = {241},
  doi       = {doi:10.1016/j.physd.2012.02.008},
  publisher = {Elsevier BV},
}

@Article{Grmela_2021,
  author    = {Grmela, Miroslav},
  journal   = {Entropy},
  title     = {Multiscale Thermodynamics},
  year      = {2021},
  issn      = {1099-4300},
  month     = jan,
  number    = {2},
  pages     = {165},
  volume    = {23},
  doi       = {https://doi.org/10.3390/e23020165},
  publisher = {MDPI AG},
}

@Article{Renger_2023,
  author    = {Renger, D. R. Michiel and Sharma, Upanshu},
  journal   = {Physical Review E},
  title     = {Untangling dissipative and Hamiltonian effects in bulk and boundary-driven systems},
  year      = {2023},
  issn      = {2470-0053},
  month     = nov,
  number    = {5},
  pages     = {054123},
  volume    = {108},
  doi       = {https://doi.org/10.1103/PhysRevE.108.054123},
  publisher = {American Physical Society (APS)},
}

@Article{Mielke_2014,
  author    = {Mielke, A. and Peletier, M. A. and Renger, D. R. M.},
  journal   = {Potential Analysis},
  title     = {On the Relation between Gradient Flows and the Large-Deviation Principle, with Applications to Markov Chains and Diffusion},
  year      = {2014},
  issn      = {1572-929X},
  month     = jun,
  number    = {4},
  pages     = {1293--1327},
  volume    = {41},
  doi       = {https://doi.org/10.1007/s11118-014-9418-5},
  publisher = {Springer Science and Business Media LLC},
}

@Article{Takhtajan_1994,
  author    = {Takhtajan, Leon},
  journal   = {Communications in Mathematical Physics},
  title     = {On foundation of the generalized Nambu mechanics},
  year      = {1994},
  issn      = {1432-0916},
  month     = feb,
  number    = {2},
  pages     = {295--315},
  volume    = {160},
  doi       = {https://doi.org/10.1007/BF02103278},
  publisher = {Springer Science and Business Media LLC},
}

@Article{Morrison_1984,
  author    = {Morrison, Philip J.},
  journal   = {Physics Letters A},
  title     = {Bracket formulation for irreversible classical fields},
  year      = {1984},
  issn      = {0375-9601},
  month     = Feb,
  number    = {8},
  pages     = {423--427},
  volume    = {100},
  doi       = {https://doi.org/10.1016/0375-9601(84)90635-2},
  publisher = {Elsevier BV},
}

@Book{degroot2013non,
  author    = {{de Groot, Sybren Ruurds} and {Mazur, Peter}},
  publisher = {Courier Corporation},
  title     = {{Non-Equilibrium Thermodynamics}},
  year      = {2013},
}

\end{document}